\documentclass[twocolumn,showpacs,preprintnumbers,amsmath,amssymb]{revtex4}

\def\beq{\begin{equation}}
\def\beql{\begin{equation}\label}
\def\eeq{\end{equation}}
\def\p{\partial}

\def\12{\frac{1}{2}}

\def\la{\label}

\usepackage{graphicx}
\usepackage{dcolumn}
\usepackage{bm}

\begin{document}

\preprint{APS/123-QED}

\title{Viscous fingering and a shape of an electronic droplet in the Quantum
Hall regime}

\author{Oded Agam}
\email{agam@amadeus.fiz.huji.ac.il}
\author{Eldad Bettelheim}
\affiliation{Racah Institute of Physics, Hebrew University, Givat
Ram, Jerusalem, Israel 91904}
\author{P.~Wiegmann}
\affiliation{James Frank Institute, Enrico Fermi Institute of the
University of Chicago, 5640 S. Ellis Ave.
Chicago IL60637}
\altaffiliation[Also at ]{Landau Institute of Theoretical Physics}
\author{A.~Zabrodin }
\affiliation{Institute of Biochemical Physics, Kosygina str. 4,
117334 Moscow, Russia}
\altaffiliation[Also at ]{ITEP, Bol. Cheremushkinskaya str. 25,
117259 Moscow, Russia}

\date{\today}

\begin{abstract}
We show that the semiclassical dynamics of an
electronic droplet confined in the plane in
a quantizing inhomogeneous magnetic field in the regime when
the electrostatic interaction is negligible is similar to viscous
(Saffman-Taylor) fingering on the interface between two fluids with
different  viscosities confined in a  Hele-Shaw cell. Both phenomena are
described by the same equations with scales
      differing by a factor of up to $10^{-9}$.
       We also report the quasiclassical wave function of the droplet in an
inhomogeneous magnetic field.
\end{abstract}


\maketitle

An important class of pattern formation on moving fronts  occurs when
diffusion, rather than convection, dominates the transport.
In these cases a front is driven with the normal velocity
proportional to the gradient of a harmonic field - a mechanism known as
D'Arcy's law or Laplacian growth (for a review see, e.g.,  \cite{review}).
Viscous  or   Saffman-Taylor fingering is one of the most
studied  instabilities of this type. It  occurs at the interface between
two incompressible fluids with different viscosities when a less viscous
fluid is injected into a more viscous one in a 2D geometry
(typically, the fluids are confined in a Hele-Shaw  cell
-- a thin gap between two  parallel  plates) or in porous media \cite{LG}.
The interface forms a pattern of growing fingers whose shape becomes
complex at high flow rates (Fig.~1).

Instabilities on diffusion driven fronts occur in different
settings ranged from geological to molecular scales. In this letter
we show  that  similar phenomena may take place in
semiconductor nanostructures. A growth of an
electronic droplet in a Quantum Hall (QH)
regime, while increasing the number of particles in the droplet,
is similar to Laplacian growth.

We first recall the Saffman-Taylor (ST) problem.
In a thin cell, the local velocity of a viscous fluid
is proportional to the gradient of pressure:
$\vec v = -\nabla p$ (the D'Arcy law). Incompressibility
implies that the pressure $p(z)$ is a harmonic function
of $z=x+iy$ with  a sink at infinity:
\beq \label{D1}
\nabla^2 p(z)=0,\quad p(z)\to
-\frac{1}{2}\log|z|,\quad |z|\to \infty
\end{equation}
If the difference between viscosities is large,
the pressure is constant  in the less
viscous fluid and, if the surface tension is
ignored, it is  also constant (set to
zero) on the interface ${\cal C}$.
Thus on  the interface
\begin{equation}
p(z)=0,\quad
      v_n =-\p_n p(z),\quad z\in {\cal C}. \label{D'Arcy}
\end{equation}

At constant flow rate, the area of the less viscous fluid
grows linearly with time $t$. We set the flow rate
such that the area equals $\pi t$.
The other parameters, viscosity and the width of the
cell, are chosen to set pressure  equal to the velocity potential.
For recent advances in the studies of fingering in channel and
radial geometries see \cite{F,mwz}.

It is convenient to rewrite Eqs.
(\ref{D1},\ref{D'Arcy}) in terms of the conformal map,
$w(z,t)$,  of the domain occupied by the more viscous
fluid to the exterior of the unit disc
$|w|\geq 1$ in such a manner that the source at $z=\infty$
is mapped to infinity.
In terms of the conformal map the pressure is
$p=-\frac{1}{2}\log| w(z,t)|$ and the
complex velocity in the viscous fluid is
$v(z)=v_x -i v_y =\frac{1}{2} \p_z \log w(z)$.
On the interface, it is
proportional to the harmonic measure:
\begin{equation}\label{v}
v_n(z,t) = \frac{1}{2}|w'(z,t)|,\;\;\;\;z\in {\cal C}.
\end{equation}
The complex velocity is conveniently written using the
Schwarz function, $S(z)$,
\beq\nonumber
\p_z\log w(z)= \p_t S(z).
\end{equation}
The Schwarz function is analytic in a domain
containing  the contour ${\cal C}$
such that $S(z)=\bar z$ on ${\cal C}$ \cite{Davis}.

In the idealized settings (\ref{D1},\ref{D'Arcy}), the
Saffman-Taylor problem confronts an obstacle. As a result of the
scale invariance, some fingers  develop
cusp-like singularities within a finite time \cite{singularities}.
A modification of the growth law which introduces a mechanism
curbing the curvature of the interface at a microscale is necessary.

The known cut-off mechanisms (e.g., surface tension,
lattice regularization, etc.) destroy the mathematical structure
  of the idealized problem and make
it difficult for  analytical analysis. It is believed,
however, that once a steady fractal pattern has been developed,
its fractal character does not depend on the mechanism
stabilizing  the singularities.

\begin{figure}
\includegraphics[width=3cm]{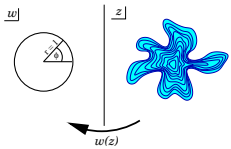}
\caption{\label{fig:wide}
A schematic illustration of a less viscous fluid domain or an electronic
droplet in a strong magnetic field. Electronic droplet is
stratified by semiclassical orbits. The domain of a more
viscous fluid (exterior of the droplet)  mapped conformally into the
exterior of a unit circle. }
\end{figure}

The purpose of this letter is twofold.
One is to show that a natural  mechanism
which introduces a scale  but captures the physical (and
the mathematical \cite{footer}) structures  of the problem is
\emph{``quantization"}.  Quantization implies that a
change of the area of the domain is quantized. This ensures
that no physical quantity can have features on a scale less
than a ``Planck constant" which cuts-off  singularities. The
Saffman-Taylor problem (\ref{D1},\ref{D'Arcy})  then
arises as a semiclassical limit, when the  scale introduced
by ``Planck's constant" tends to zero.

Our second goal is to show that quantized
Saffman-Taylor problem  describes an electronic
droplet in a special QH-regime where electrostatic forces
are  weak. This correspondence
suggests that the edge of the QH-droplet  may develop unstable
features similar to the fingers observed in Hele-Shaw
cell (a fingering pattern is illustrated in Fig.~1).  A spatial
resolution of nanoscale structures in GaAs by recently developed
scanning techniques might be apt for a search for such
features \cite{QH}.

We  will study a shape of a large electronic droplet  on the fully
occupied lowest Landau level of a quantizing magnetic field.
The magnetic field  is assumed to be  nonuniform in the area  away from
the droplet.  We show that Aharonov-Bohm forces,  associated with the
nonuniform part of the magnetic field shape the edge of the droplet in
a manner similar to a fingering interface driven by a Laplacian
field.

In order to present our argument we shall
neglect the interactions among the
electrons and assume that the external electrostatic
potential is zero. Under these conditions we will
show that the semiclassical dynamics of the QH droplet is
governed by the same equations of viscous fingering scaled to
a nanometer scale. By the semiclassical limit we mean a
large number of electrons $N$, small magnetic length but a
finite area of the droplet:
\beq\nonumber
N\to \infty,\quad \ell \equiv\sqrt{
\hbar c/eB_0}\to 0,\quad
t=2 \ell^2 N =\mbox{finite}.
\end{equation}
The droplet grows when its quantized area changes
by a quantum, $\pi t\to\pi t+2\pi \ell^2$, as
an electron is added ($N\to N+1$) or by changing
the magnetic length.

Prior to computations,  we would
like to give an insight why electronic droplet
in magnetic field may behave similarly to viscous fingering.
Equations (\ref{D1}, \ref{D'Arcy}) imply that
the harmonic moments of the viscous fluid domain,
\begin{equation}\nonumber
t_k= -\frac{1}{\pi k} \int_{\rm visc. fluid}\!
{z^{-k}}{d^2z},\quad
k=1,\,2,\ldots
\end{equation}
do not change in time \cite{R}. They are initial data of evolution.
Indeed,
$\frac{ d}{dt}t_k
=\frac{1}{\pi k}\oint_{{\cal C}}
z^{-k}\partial_n p(z)|dz|=
\frac{1}{2\pi i k}\oint_{{\cal C}}
z^{-k}\partial_{z}\log w(z) dz =0$
since $\p_z \log w(z) $ is  analytic in the viscous domain.
Conservation of the harmonic moments is an
equivalent formulation of D'Arcy's law (\ref{D'Arcy}).

As is shown below, the harmonic moments of the
QH-droplet do not depend on the number of particles in
the droplet. The moments feature the inhomogeneity of the magnetic field.
A proof of this assertion  would yield D'Arcy's law for the QH droplet.

Our proof consists of two steps. First we
construct the wave functions of the QH-states in a weakly nonuniform
magnetic field and introduce the $\tau$-function - a generating
function of the matrix elements.
Then we determine  semiclassical states - the
orbits (supports of the wave functions),
and study an evolution of the orbit as the area of
the droplet is increased. We will consider the droplets with smooth shape,
leaving analysis of the
finger-tip  singularities for  further publications. The size of the
letter forces us to rely on some formulas on the evolution of conformal
maps proved in \cite{kkmwz}.

Let us first recall the concept of a QH-droplet (see e.g.,
\cite{Laughlin,Cappelli}). Consider spin polarized electrons
\cite{comment} on a plane in the lowest level of a
quantizing nonuniform magnetic field, directed perpendicular to the
plane, $B(x,y)>0$:
$$
      H=\frac{1}{2m}\left ( (-i\hbar \vec\nabla -\vec A)^2
      - \hbar B \right ).
$$
The lowest  level of the Pauli Hamiltonian  is degenerate
even for a nonuniform field. The  degeneracy equals the
integer part of the total magnetic flux $\Phi= \int dx dy B$
in units of flux quanta, $\Phi_0= 2 \pi\hbar$ (we set $e=c=1$).
The orthonormalized degenerate states, written in transversal gauge, are
$\psi_n(x,y) =P_{n}(z)\exp{\frac{W}{\hbar}},$ where the potential
$W(x,y)$ obeys the equation $B=-\nabla^2 W$, and
$P_n(z)$ are holomorphic polynomials  of a degree $n$, which does not
exceed the degeneracy of the level
\cite{grstate}.

We will consider the following arrangement: A strong uniform magnetic
field $B_0>0$ is situated in a large disk of radius $R_0$; The disk
is surrounded by a large annulus $R_0 <|z|< R_1$ with a  magnetic
field $B_1<0$ directed  opposite to $B_0$, such that the total magnetic
flux $\Phi$ of the disk, $|z|<R_1$, is $N \Phi_0$. The magnetic field
outside the disk  $|z|<R_1$ vanishes. The  disk is connected  through a
tunneling barrier to a large capacitor that maintain a uniform small
positive chemical potential slightly above the zero energy of the
lowest Landau level.

In this arrangement a circular droplet of $N$ electrons
is trapped at the center of the disk $|z|<R_0$. We choose
the magnetic field $B_1$ such that its radius $\ell\sqrt{2N}$
is much smaller than the radius of the disk $R_0$.

Next we assume that a weakly nonuniform magnetic field
$\delta B$ is placed inside the disk $|z|<R_0$ but well away from the
droplet. The nonuniform magnetic field does not change
the total flux $\int\delta B dx dy=0$. The droplet grows when
$B_1$ is adiabatically increased, keeping
$B_0$ and $\delta B$ fixed. Then the degeneracy of the Landau level
and, consequently, the size of the droplet increase.

The wave function of the droplet is the
Jastrow function
\begin{equation}\label{VdM}
\Psi(z_1,\dots,z_N)=\frac{1}{\sqrt{N!\tau_{N}}} \Delta (z)
e^{\frac{1}{ \hbar}\sum_nW(z_n) },
\end{equation}
where
\begin{eqnarray}
\Delta (z) =\prod_{n<m<N}(z_n-z_m) = \sqrt{ \tau_N} \det \big
(P_{n}(z_m)\big)\big |_{n,m< N}
\nonumber
\end{eqnarray}
is the Van der Monde determinant, and the normalization
factor  - the $\tau$-function-
\beq\la{tau}
\tau_N= \frac{1}{N!}\int |\Delta(\xi)|^2\prod_ne^{\frac{2}{\hbar}  W(\xi_n
)}d^2\xi_n\,.
\end{equation}
depends on magnetic field.

In terms of the orthonormal one-particle states, 
$\psi_n(z)$, the density $\rho_N(z)=N\int
|\Psi(z,\xi_1,\dots,\xi_{N-1})|^2d^2\xi$ reads
\begin{eqnarray}
\rho_{N+1}(z)-\rho_{N}(z)= |\psi_{N}(z)|^2.\nonumber
\end{eqnarray}
This formula defines a growth process. One-particle
state gives the probability of adding an extra particle
to a point $z$ of the droplet.

The integral in (\ref{tau})  converges in the area of the
droplet, where magnetic field  is uniform. In this area the
annulus, $R_0<|z|<R_1$,  does not contribute to
the potential $W$, and the potential $V$
of the nonuniform part of the magnetic field
$\delta B=-\nabla^2 V$ is a harmonic function
\begin{equation}\nonumber
V(z)=Re \sum_{k\geq 1} t_k z^k,\;\;\;\;\;\;\;\, t_k=\frac{1}{2\pi k
}\int
\delta B(z)z^{-k} d^2z.
\end{equation}
The parameters $t_k$ are, now, the  harmonic
moments of the deformed part of the magnetic field.
Summing up, we have $W(z)=-{|z|^2}/{4\ell^2}+V(z)$.

The one-particle state $\psi_N (z)$
can be expressed in terms of the $\tau$-function
(further below we set
$2\ell^2=\hbar$)
\begin{equation}\label{Pnn}
\psi_N (z)=
\frac{e^{-\frac{1}{\hbar}(\frac{|z|^2}{2}-{V(z)})}}
{\sqrt{(N+1)\tau_{N+1}\tau_N}}z^N
e^{-{\hbar} D(z)}\tau_N,
\end{equation}
where
$D(z)=\sum_{k\geq 1}\frac{z^{-k}}{k}\frac{\p}{\p
t_k}$ is a generator of the deformations of the magnetic field.

The next step is a semiclassical analysis of the wave function,
$\psi_N (z)$. In this limit the droplet has a sharp
boundary (we assume that the boundary is smooth), and the
semiclassical states (with large $N$) are localized at orbits of
width of the order of $\ell$. Since the magnetic field in the area
of the droplet is assumed to be uniform, the density is also
uniform inside the droplet. The area of the droplet is
quantized as $\pi t=2\pi N\ell^2$.

As $\hbar\to 0$ with $t=\hbar N$
fixed, the integral (\ref{tau}) is saturated by its saddle
point. At the saddle point, the $\xi_n$ are uniformly
distributed in a domain determined by the area $\pi \hbar N$
and the harmonic moments $t_k$  \cite{kkmwz}.
We will see that  this domain determine the shape of the
droplet. At the saddle point
the $\tau$-function (\ref{tau}) tends to the classical
  $\tau$-function
$F=\lim_{\hbar\to 0}\hbar^2\log\tau_N$,
studied in \cite{kkmwz}:
\beq\nonumber
F(t;t_1,\dots)
=-\frac{1}{\pi^2}
\int\int_{droplet} \log \left |
\frac{1}{z}-\frac{1}{z'}
\right |d^2 z d^2 z'.
\end{equation}

On expanding (\ref{Pnn}) in $\hbar$, and
defining
\beq\nonumber
\Omega(z)=\sum_{k\leq 1}t_k z^k+t\log z-\left(\frac{1}{2}
\partial_t+ D(z)\right)F,
\end{equation}
and
$${\cal A}(z,\bar z)=\frac{1}{2}|z|^2-Re\,\Omega(z),$$
one can write  the amplitude of a semiclassical state, in
terms of variations of the classical $\tau$-function:
\begin{equation}
|\psi_N (z)|^2 \simeq
\frac{
e^{-\frac{1}{2}\left(\partial^2_t -2 Re\, D^2(z)\right )F}}
{\sqrt{2\pi^3 \hbar}}\,
e^{-\frac{2}{\hbar}{\cal A}(z,\bar z)},\;\;t=\hbar N.
\label{psi}
\end{equation}
Properties of the functions
$\Omega(z)$ and $D^2(z)F$ clarify the
meaning of a semiclassical state. We list them below.
The function ${\cal A} (z)$  is the imaginary part of the semiclassical
action.  It was proved in \cite{mwz,kkmwz}
that ${\cal A}$
\begin{itemize}
\item [-]
      reaches its minimum on a
closed contour (an orbit) which bounds
the domain with harmonic  moments
$t_k$ and the area  $\pi t$, and  everywhere on  the orbit
the action vanishes:
\beq\nonumber
{\cal A}(z)=0,\quad Re\,
\Omega(z)=\frac{1}{2}|z|^2,\quad  z\in\mbox{orbit}.
\end{equation}
\item [-]
The first derivative of the action
normal to this contour also vanishes on the orbit:
\beq\label{S}
\p_z{\cal A}(z)=0,
\quad\quad\partial_z\Omega(z)=\bar z,\quad z\in\mbox{orbit}.
\end{equation}
\item [-]
If the area of the contour increases, while $t_k$ remain fixed,
the action changes as
\beq\nonumber
\partial_t{\cal A}=-\partial_t Re\,\Omega(z,t)=- \log |w(z,t)|.
\end{equation}
\end{itemize}
Eq.(\ref{S}) implies that $S(z)\equiv\partial_z\Omega(z)$ is the
Schwarz function
of the contour.

The  the action ${\cal A}$ remains positive
in the vicinity of the contour and everywhere
in the exterior domain. Its second variation normal to the contour reads
\beq\nonumber
{\cal A}(z+\delta_n z)=|\delta_n z|^2+\ldots\,,
\quad z\in {\cal C}
\end{equation}
where $\delta_n z$ is a normal deviation from  a point $z$ on the contour
(we use the relation
$|\delta_n z|^2=-S'(z)(\delta_n z)^2$).

A semiclassical state is localized at the
minimum of ${\cal A}(z)$, where the amplitude  has a sharp maximum.
A contour where the action vanishes is the
orbit.
We conclude that all orbits have the same
harmonic moments $t_k$ and differ by the area.
This  implies the D'Arcy law.

The second variation of the classical $\tau$-function
was also computed in Ref.~\cite{kkmwz}:
\beq\nonumber
\log \frac{w(z)-w(z')}{z-z'}=
D(z)D(z')F-\frac{1}{2}\partial _{t}^{2}F.
\end{equation}
Merging the points $z$ and $z'$,
we find that the value of the first factor
in (\ref{psi})
is just the
harmonic measure $|w'(z)|$.

      Summing up, the
amplitude of the semiclassical  wave function in a weakly inhomogeneous
magnetic field  is
\beq\nonumber
|\psi_N(z)|^2\simeq
{\frac{|w'(z)|}{\pi\sqrt{2\pi\hbar}}}
e^{-\frac{2}{\hbar}{\cal A}(z,\bar z)},\quad
t=N{\hbar}.
\end{equation}
      The factor in front of the exponent ensures the correct normalization:
$|\psi_N|^2 ds=d\phi
(2\pi^3\hbar)^{-1/2}e^{-\frac{2}{\hbar}|\delta_nz|^2}$
(here $ds$ is an element
of the length of the orbit,  and
$d\phi$ is an element of the angle on the unit circle).
The classical
probability distribution is then
$|\psi_N|^2 \simeq\frac{1}{2\pi}|w'(z)|\delta_{\cal C}(z)$,
where the  $\delta$-function is localized on the orbit.

When a new particle is added to the droplet,
the edge advances. The
velocity $v_n$ of the advance is defined as
$\partial_t\rho(z)=v_n\p_n \rho(z)$. The normal
derivative of the density is $\p_n\rho
=\frac{1}{\pi\hbar}\delta_{\cal C}(z)$, and a change of the
density is
$\hbar\partial_t\rho(z)\simeq
\rho_{N+1}-\rho_{N}=|\psi_N|^2\simeq
\frac{1}{2\pi}|w'(z)|\delta_{\cal C}(z).
$
This prompts
      D'Arcy's law (\ref{v}).

It is interesting to compare dimensionless viscosity of liquids
used in  viscous fingering experiments
\cite{viscous_exper}, and a "viscous" effect of quantum interference of
at the first Landau level. The  parameter of the
dimension of length controlling viscous fingers in fluids -- an effective
capillary number -- is $\frac{2\pi}{q}\frac{b^2}{12\eta}\sigma$,
where $q$ is the flow
rate, $b$ is the thickness of the cell, $\eta$ is the viscosity and
$\sigma$ is the surface tension. In recent experiments
\cite{viscous_exper}, this number stays in high
hundred of nanometers,
but can be easily decreased by increasing the flow rate. This length
is to be compared with  the magnetic length $\ell$.
At magnetic field about 2T it is
about  50 nm. At these numbers, the capillary effects
are less important  in electronic liquids than
in classical fluids, but remarkably stay in the same range.
Under conditions reducing electrostatic effects,
fingering  in  electronic droplets
may be even more dramatic than in viscous fluids.
Semiconductor devices imitating a channel geometry
of the original Saffman-Taylor experiment
\cite{LG} may facilitate fingering instability.

We acknowledge useful dis\-cus\-sions with
A. Bo\-yar\-sky, B. Da\-vi\-do\-vich, A. Gor\-sky,
L. Le\-vi\-tov, A. Mar\-sha\-kov,
M. Mi\-ne\-ev-\-Wein\-stein, A. Cappelli,
V. Ka\-za\-kov,
I. Kos\-tov, L. Ka\-da\-noff,
W. Kang, I. Pro\-cac\-cia, O. Ru\-chay\-skiy, N. Zhi\-te\-nev.
P.W. and A.Z.
were supported by grants NSF DMR 9971332 and MRSEC NSF DMR 9808595.
P. W. thanks C.Gruber for the hospitality in EPFL, where the paper was
completed. The work of A. Z. was partially supported by grants
CRDF RP1-2102, INTAS-99-0590 and RFBR 00-02-16477.
This research was supported in part by  THE ISRAEL
SCIENCE FOUNDATION founded by The Israel Academy of Science
and Humanities, and by Grant No.~9800065 from the USA-Israel
Binational Science Foundation (BSF).


\begin{thebibliography}{10}
\bibitem{footer} Elsewhere we show that quantization retains
the integrable structure of the ``classical" problem found
in Ref. \cite{mwz}. In particular, the function
(\ref{tau}) appears to be the $\tau$-function of Toda
lattice hierarchy.
\bibitem{comment} We consider
relativistic electrons only for the sake of simplifying the
derivation of  the semiclassical wave function.
\bibitem{C} Despite of the difference,  both phenomena have
      similar algebraic structures. The operator content of the
Quantum Hall effect forms the quantum $W_{1+\infty}$ algebra,
while the Laplacian Growth  can be viewed as an action of the
classical $w_{1+\infty}$ algebra.
The algebras reflect the incompressibility of the quantum and
      the classical liquids respectively \cite{Cappelli}.
\end{thebibliography}
\end{document}